\pdfoutput=1

\documentclass[11pt]{article}

\usepackage[preprint]{acl}

\usepackage{times}
\usepackage{latexsym}

\usepackage[T1]{fontenc}

\usepackage[utf8]{inputenc}

\usepackage{microtype}

\usepackage{inconsolata}

\usepackage{graphicx}

\usepackage{subcaption}

\usepackage{booktabs}

\title{Extracting Information About Publication Venues Using Citation-Informed Transformers}

\author{Brian D. Zimmerman \\
  University of Waterloo \\
  \texttt{bdzimmer@uwaterloo.ca} \\\And
  Josh Folkins \\
  University of Waterloo \\
  \texttt{jfolkins@uwaterloo.ca} \\ \\\And
  Olga Vechtomova \\
  University of Waterloo \\
  \texttt{ovechtomova@uwaterloo.ca} \\}

\begin{document}
\maketitle
\begin{abstract}
Scientific document embeddings contain a variety of rich features which can be harnessed for downstream tasks such as recommendation, ranking, and clustering. We explore which tangible insights can be drawn from scientific document embeddings to understand trends in computer science research featured across nine well-known venues. We collect approximately 60,000 scientific documents published between 2015 and 2023 and analyze their embeddings, which we produce with the SPECTER pre-trained language model. In particular, we examine whether similarity between two venues can be measured using the embeddings of the scientific documents they admit for publication. Our findings indicate that some venues within computer science are indistinguishable when only considering the distributions of their document embeddings. We additionally examine whether any two venues are becoming increasingly similar over time and identify a trend of convergence within some venues in our analysis. We discuss the implications of these results and the potential impact on new scientific contributions.
\end{abstract}

\section{Introduction}
The recent increase in scientific participation has led to an overwhelming increase in scientific document publication. In the most active areas of research, a call for papers can receive thousands of submissions\footnote{https://github.com/lixin4ever/Conference-Acceptance-Rate}. We describe this situation as an ``information bottleneck'', where researchers have more information available to them than they can reasonably ingest.

Services such as the Semantic Scholar literature graph~\citep{ammar2018construction} rely on embeddings produced by SPECTER-2~\citep{singh2022scirepeval}, a citation-informed transformer encoder, to track the relationship between scientific documents and construct literature feeds for researchers that align with their interests. Scientific document embeddings are highly informative due to the incorporation of document contents and citation data during encoder pre-training. Embeddings produced by language models that incorporate citation data ~\citep{cohan2020specter, singh2022scirepeval} have been very successful in recommendation, ranking, and clustering applications. Although citation-informed encoders provide excellent empirical results, they remain difficult to interpret due to a lack of a clear framework for doing so~\citep{mittelstadt2019explaining}.

\begin{table}[]
    \centering
    \small
    \begin{tabular}{c|c|c}
        \toprule
        Venue & Category & \# Papers\\
        \midrule
        NeurIPS & Artificial Intelligence & 13884 \\
        AAAI & Artificial Intelligence & 11523 \\
        CHI & Human-Computer Interaction & 10811 \\
        ACL & Natural Language Processing & 8133 \\
        EMNLP & Natural Language Processing & 7453 \\
        NAACL & Natural Language Processing & 3968 \\
        USS & Privacy and Security & 1764 \\
        OOPSLA & Programming Languages & 682 \\
        POPL & Programming Languages & 531 \\
        \bottomrule
    \end{tabular}
    \caption{Computer science venues considered for our analysis. All articles were published between 2015 and 2023.}
    \label{tab:hvenues}
\end{table}

We use document embeddings produced by SPECTER~\cite{cohan2020specter} to analyze the information bottleneck through the lens of computer science publication venues. By combining encoded documents with their publication metadata, we extract and interpret trends across nine well-known computer science venues to answer the following research questions:

\begin{enumerate}
    \item {
        Can the similarity of two venues be measured by comparing the respective documents they accept for publication?
    }
    \item {
        Are scientific documents published in computer science venues becoming increasingly similar?
    }
\end{enumerate}

We additionally release Mizzium Library, a free visualization tool for identifying interdisciplinary research opportunities. We include all of the documents analyzed in this work as part of our initial catalogue in Mizzium Library, which can be explored at \url{https://mizlib.com}.

\begin{figure}[t]
    \centering
    \includegraphics[width=0.45\textwidth]{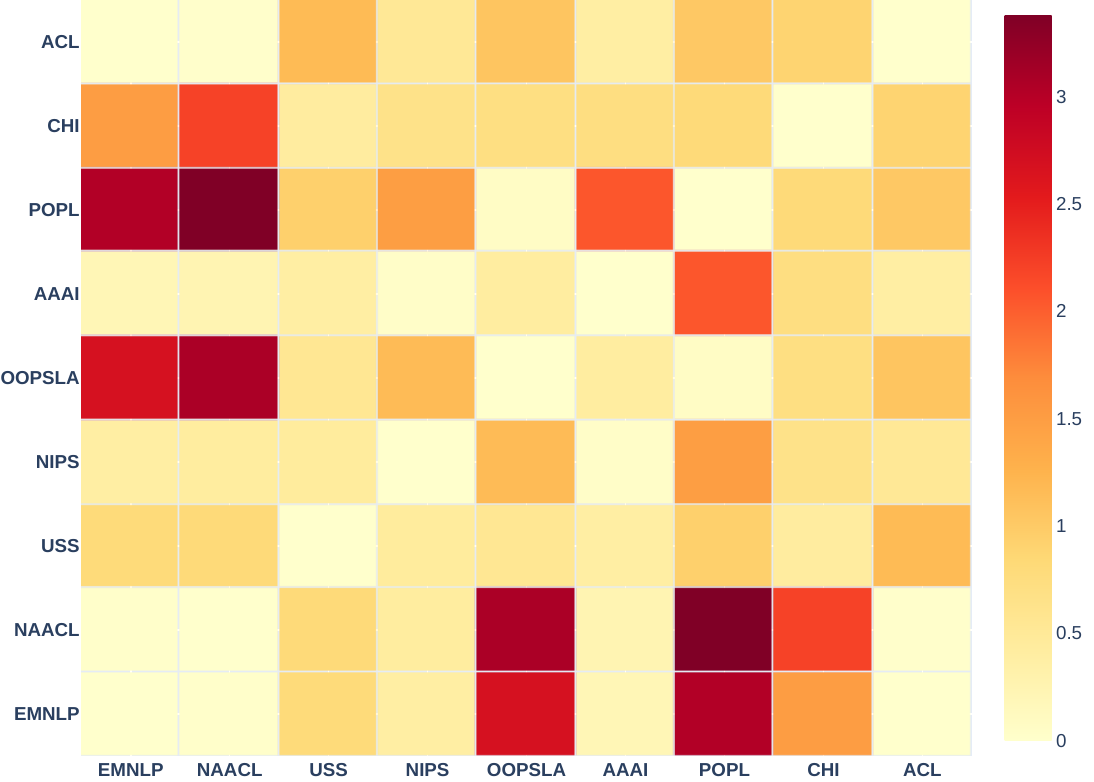}
    \caption{Variance-weighted Kullback-Leibler (KL) divergence between each venue. Lower values indicate two venues are more similar.}
    \label{fig:kld_square}
\end{figure}

\section{Experimental Setup}
For our experiments, we select nine well-known venues in computer science research on the basis of their reputations for participation in interdisciplinary research. We assign each venue a general category in Table~\ref{tab:hvenues}.

\subsection{Dataset}
After selecting the venues, we use the DBLP API to collect titles and other unique identifier metadata by venue. Using the unique identifiers, we then collect abstracts from the bulk retrieval endpoint of the Semantic Scholar API. We opt to use DBLP for our venue-based retrieval because we encountered several mislabeled articles in Semantic Scholar's database.

We use the SPECTER citation-informed encoder~\citep{cohan2020specter} with pretrained weights retrieved from HuggingFace to encode each document. Although SPECTER can encoder titles without abstracts, we remove entries for which abstract contents could not be retrieved from Semantic Scholar. In total, we collect 58,749 documents accepted for publication within one of the nine venues between 2015 and 2023.

\subsection{Experiments}
SPECTER produces 768 dimensional embeddings by jointly encoding a title and an abstract. As a preprocessing step, we use principal component analysis (PCA) to extract more manageable 64-dimensional feature vectors while preserving 82.5\% of the variance of the original distribution.

\begin{equation}
    D_{KL}(P \parallel Q) = \int_{-\infty}^{\infty} p(x) \log \frac{p(x)}{q(x)} \, dx
    \label{eq:kld}
\end{equation}

\subsubsection{Variance-Weighted KL Divergence as a Similarity Metric}
To evaluate the similarity between two venues, we measure the alignment between the distributions of their feature vectors. For each combination of venues, we compute the Kullback-Leibler (KL) divergence between them across each individual feature (Equation~\ref{eq:kld}). We consider the aggregate KL divergence across each feature to be the measure of similarity between two venues, with lower values indicating that the distributions of the two venues are more similar.

Features of the 64-dimensional PCA embeddings are ordered by how much variance they encode from the original, high-dimensional distribution. The first feature of each vector encodes the highest amount of variance and should be treated with a higher amount of importance. We normalize the ratio of variance encoded into each feature and compute the dot product with the KL divergence scores for each feature, resulting in a variance-weighted sum. This process emphasizes differences in higher variance features while simultaneously minimizing differences in lower variance features.

\subsubsection{Evaluating Variance Year Over Year}
We group each document by year in order to evaluate venue similarity over time. For each group, we again evaluate each venue against each other venue using variance-weighted KL divergence to observe whether the distributions of documents attributed to two venues are moving closer together or further apart.

For each combination of venues, we train a logistic regression classifier on documents within each yearly group. Two collections of documents sampled from similar distributions will be more difficult for a logistic regression classifier to accurately assign a label to. The intuition in using logistic regression is that any combination of venues in which the regression model exhibits low validation accuracy likely have some thematic or categorical overlap.

\begin{figure}[t]
    \centering
    \begin{subfigure}[b]{0.45\textwidth}
        \centering
        \includegraphics[width=\textwidth]{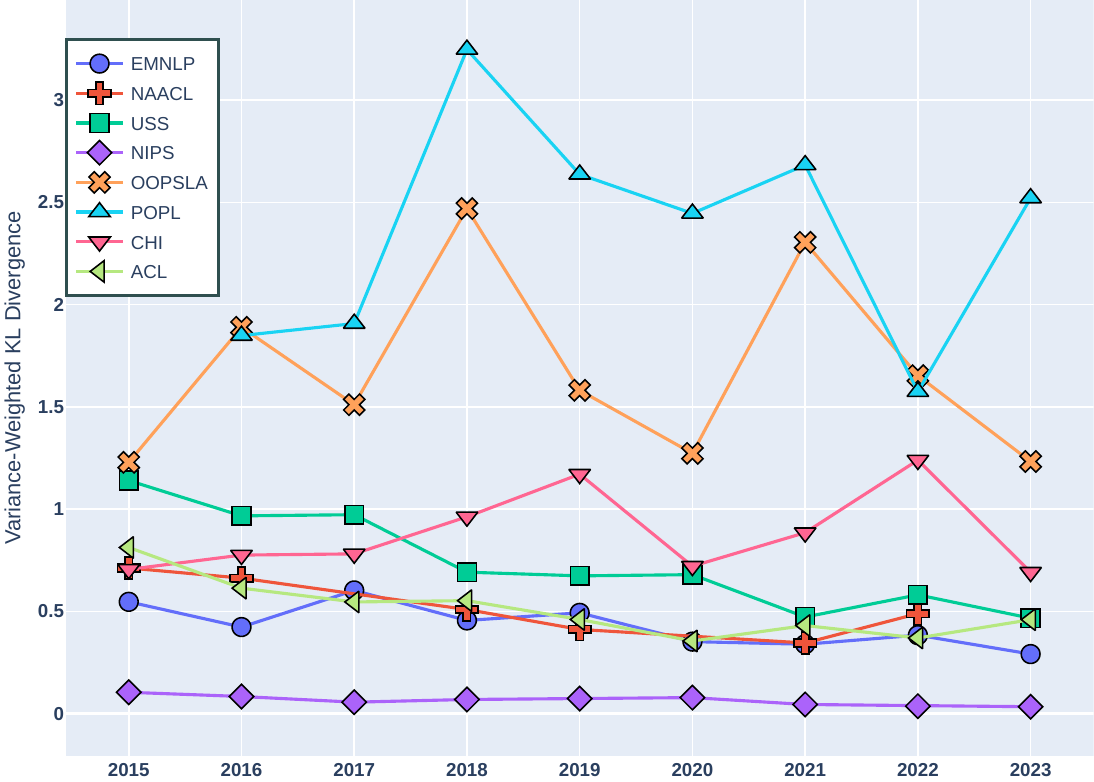}
        \caption{AAAI (Variance-weighted KL divergence)}
        \label{fig:sub1}
    \end{subfigure}
    \hfill
    \begin{subfigure}[b]{0.45\textwidth}
        \centering
        \includegraphics[width=\textwidth]{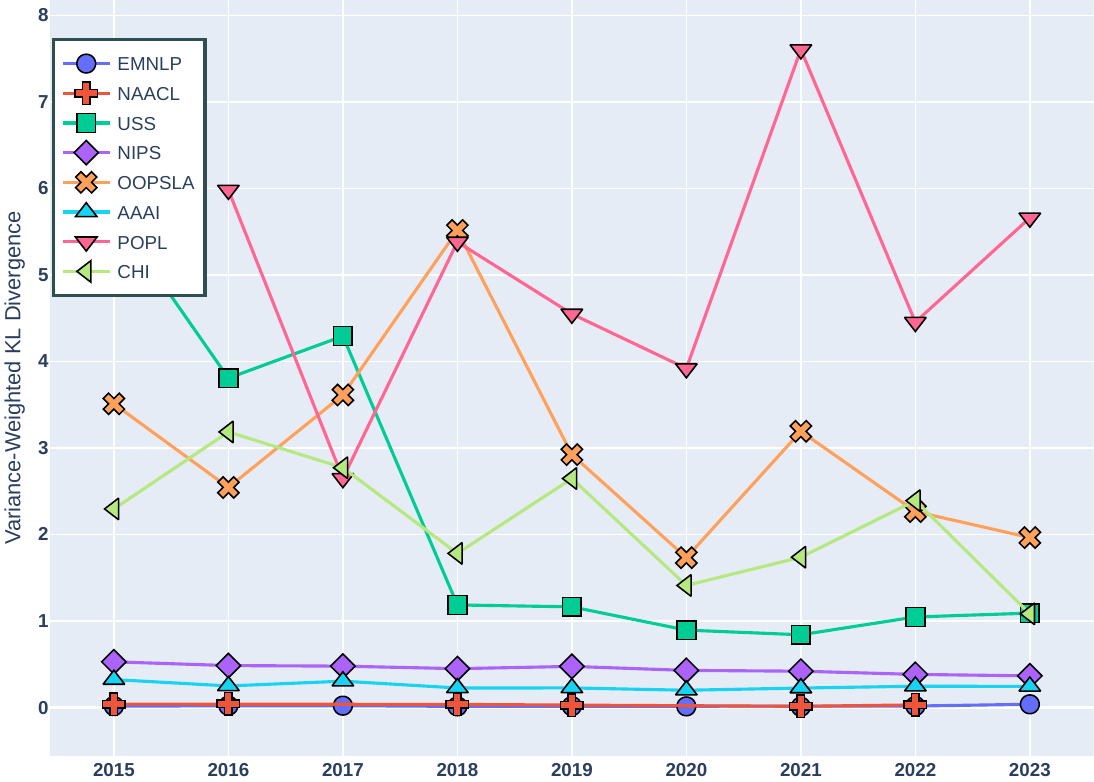}
        \caption{ACL (Variance-weighted KL divergence)}
        \label{fig:sub2}
    \end{subfigure}
    \caption{We measure the variance-weighted KL divergence over time for AAAI and ACL. We perform within venue normalization for readability, as some venues were divergent by an order of magnitude higher than others.}
    \label{fig:kld_yearly}
\end{figure}

\begin{figure}[t]
    \centering
    \begin{subfigure}[b]{0.45\textwidth}
        \centering
        \includegraphics[width=\textwidth]{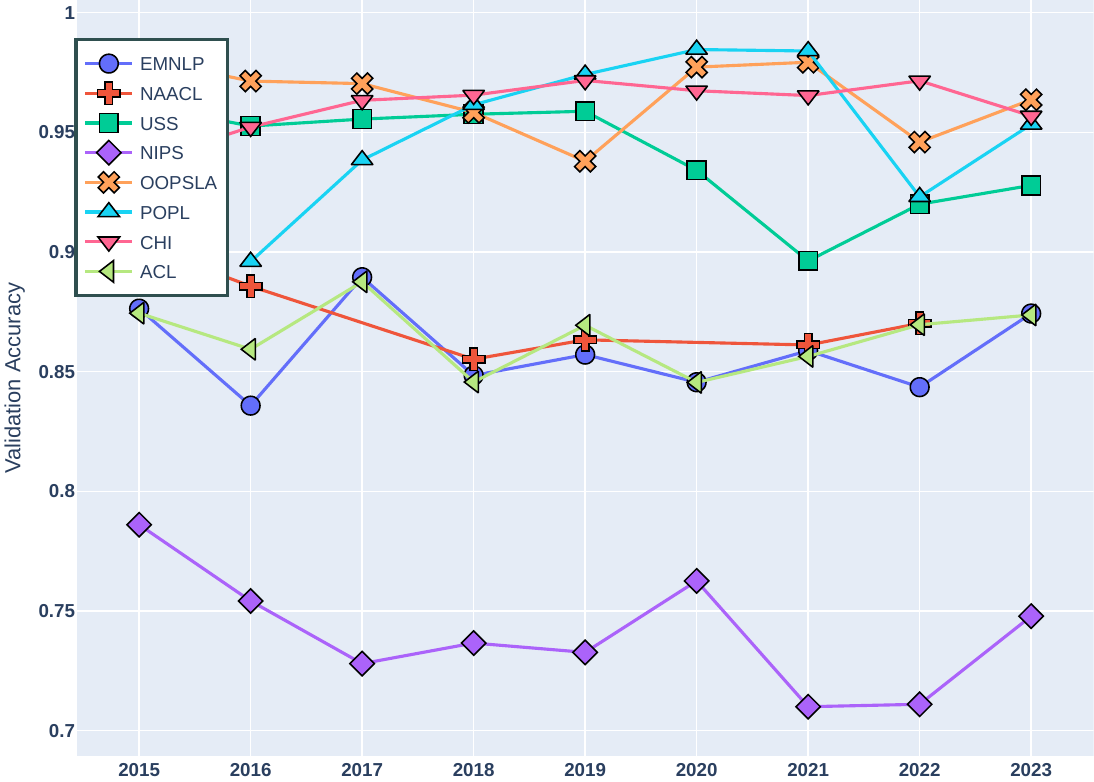}
        \caption{AAAI (logistic regression)}
        \label{fig:sub1}
    \end{subfigure}
    \hfill
    \begin{subfigure}[b]{0.45\textwidth}
        \centering
        \includegraphics[width=\textwidth]{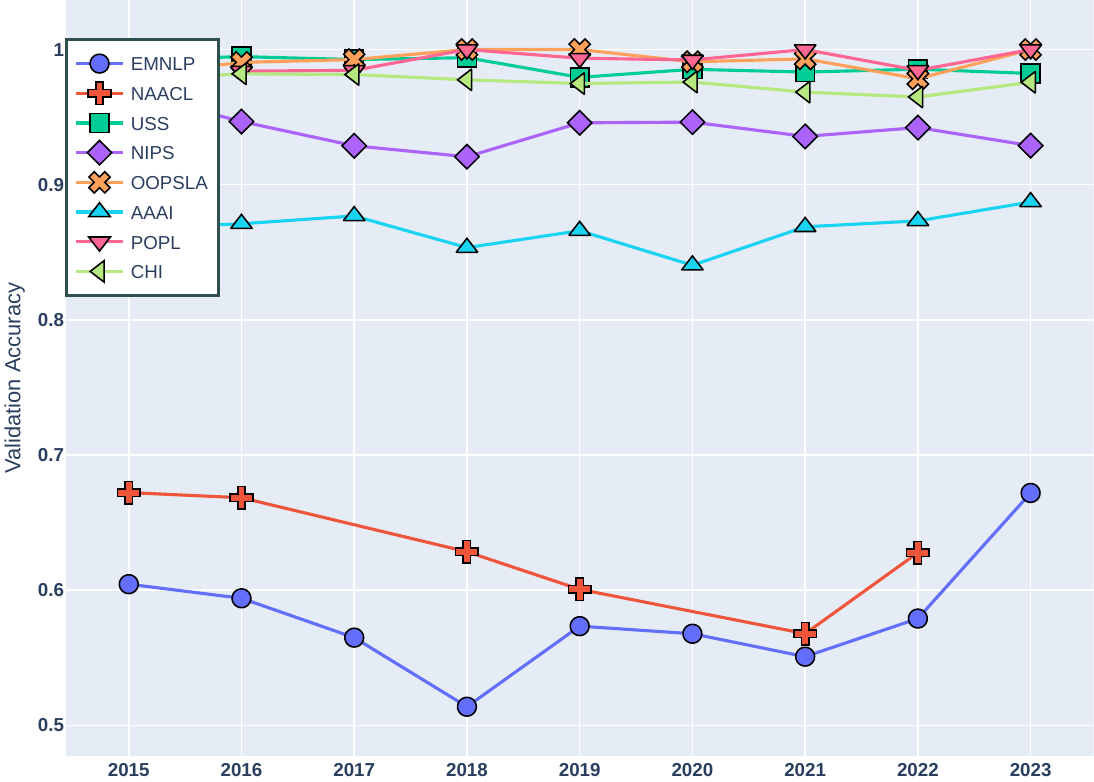}
        \caption{ACL (logistic regression)}
        \label{fig:sub2}
    \end{subfigure}
    \caption{For each combination of venues, we group their documents by year and train a logistic regression classifier on each group. We report the validation accuracy above for AAAI and ACL as they compare with each other venue. We expect venues within the same category to be more difficult for a classifier to discern, resulting in lower validation accuracy.}
    \label{fig:logr_yearly}
\end{figure}

\begin{figure*}[h!]
    \centering
    \includegraphics[width=0.98\textwidth]{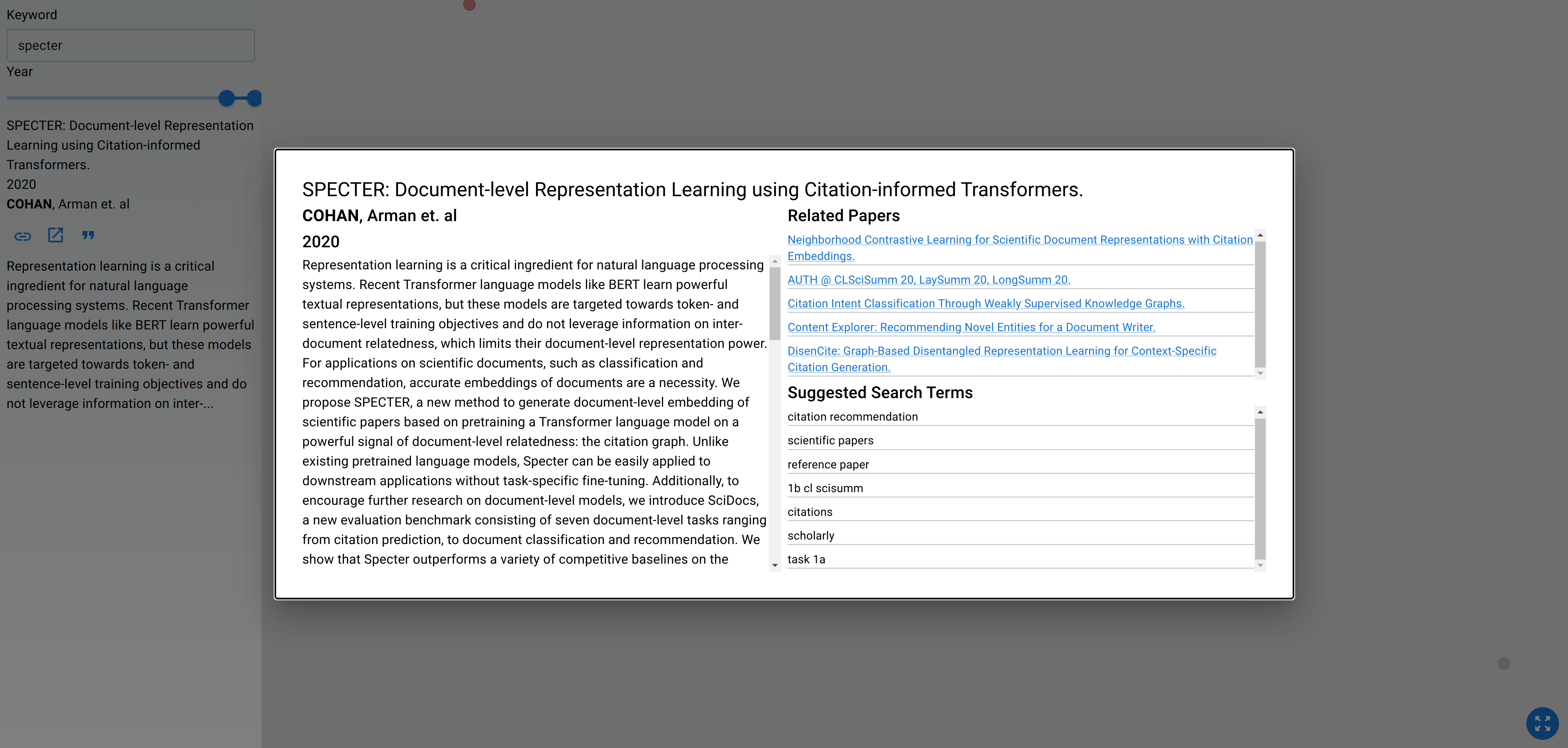}
    \caption{A screenshot of the Mizzium Library interface, with search terms extracted from related documents.}
    \label{fig:mizlib}
\end{figure*}

\section{Results and Discussion}
We first discuss the overall similarity between venues in the measured time period before discussing the utility of our metric year over year.

\subsection{Overall Similarity Between Venues}
We report the variance-weighted KL divergence in Figure~\ref{fig:kld_square} and immediately observe a pattern between venues. Each of ACL, NAACL, and EMNLP have negligible distance between the distributions of their respective features, indicating that their respective document pools are co-located in the PCA feature space. Similarly, the ``programming language'' venues OOPSLA and POPL are co-located with one another in the embedding space but distinct from each other venue. The venues CHI, AAAI, and NeurIPS have general themes, but their large publication volume invites a wide variety of interdisciplinary research.

\subsection{Year Over Year Similarity}
The year over year variance-weighted KL divergence between each combination of venues reveals that document distributions between some pairs of venues are becoming closer to one another. 

In Figure~\ref{fig:kld_yearly}, we examine AAAI and ACL against each other venue with respect to their variance-weighted KL divergence. Documents accepted to AAAI appear to be consistently distinct from those of each programming language conference. However, AAAI is steadly moving closer to each NLP conference included in our analysis, a phenomenon that may reflect recent trends around large language models. The results of the logistic regression classifier in Figure~\ref{fig:logr_yearly} appear to consolidate this effect. A similar trend can be observed between ACL and interdisciplinary conferences such as USS and CHI. Interdisciplinary topics within language modelling such as enforcing privacy guarantees on language models or human-in-the-loop systems could be hypothetically be admitted for publication at any of these venues.

\subsection{Measuring the Health of Scientific Venues}
Venues are accepting more submissions year over year as scientific participation increases. Our analysis shows that some venues are converging, indicating a potential loss of thematic identity. As more researchers begin to rely on Large Language Models (LLMs) as writing assistants, writing style between venues may also become more similar. Establishing a framework for monitoring the health of scientific venues would help research communities determine when inflated venues should split into two or consider shifts in their organization. 








\subsection{Mizzium Library -- A Tool for Identifying Interdisciplinary Research Opportunities}
One motivation for this work was to build a tool that could assist in identifying interdisciplinary research opportunities. Mizzium Library (Figure~\ref{fig:mizlib}) is a graphical catalogue of scientific papers, encouraging exploratory surveys. Papers are presented as points in a scatter plot, which users can click on to view more details such as related papers or suggested search terms. Suggested search terms are extracted as common spans of text in the pool of related documents. Users can follow suggested search terms as breadcrumbs, exploring documents containing concepts which they may have been unfamiliar with prior. Our intention for Mizzium Library is to continue its development by adding new features and ingesting data from venues outside of computer science.

\section{Related Work}
To our knowledge, this work is the first to explore the use of embeddings produced by a citation-informed transformer to describe relationships between publication venues. 

~\citet{hacohen2013classifying} rely on feature engineering, hand-selecting 103 features to train a supervised classifier. ~\citet{garcia2016analyzing} rely on features of authorship and citation networks to produce similarities between different venues. By contrast, we rely on citation-informed transformers, which encode titles and abstracts as natural language to produce document representations without expensive feature extraction.

\section{Conclusion}
In this paper, we measure the similarity between nine well-known computer science venues by analyzing the distributions of their encoded documents. We observe clear patterns between venue themes and the alignment of their distributions, as measured with our proposed variance-weighted KL divergence. Additionally, we find that the distributions between some venues are converging over time. Our data is catalogued in Mizzium Library, a free online tool for identifying interdisciplinary research opportunities.

\bibliography{custom}
\end{document}